# Lossless Brownian information engine

Govind Paneru,[1] Dong Yun Lee,[1,†] Tsvi Tlusty,[1,2] and Hyuk Kyu Pak[1,2*]

[1]Center for Soft and Living Matter, Institute for Basic Science (IBS), Ulsan 44919, Republic of Korea

[2]Department of Physics, Ulsan National Institute of Science and Technology (UNIST), Ulsan 44919, Republic of Korea

**We report on a lossless information engine that converts nearly all available information from an error-free feedback protocol into mechanical work. Combining high-precision detection at resolution of 1 nm with ultrafast feedback control, the engine is tuned to extract the maximum work from information on the position of a Brownian particle. We show that the work produced by the engine achieves a bound set by a generalized second law of thermodynamics, demonstrating for the first time the sharpness of this bound. We validate a generalized Jarzynski equality for error-free feedback-controlled information engines.**

Understanding the interplay between information and thermodynamics is a fundamental challenge of nonequilibrium physics, in particular in systems of active and living matter that self-organize into information-rich homeostatic ensembles. The question emerged with Maxwell's demon who, by measuring the velocity of gas molecules, was able to sort them into fast and slow ones, thus decreasing the entropy and apparently violating the second law of thermodynamics [1]. A series of works, starting from Szilard's engine [2] through Landauer [3], Bennett [4] and others, elucidated the link between information gathered by the demon and thermodynamic entropy, thereby resolving the apparent paradox. That the demon can extract work from information has been known since these seminal papers, but recent breakthroughs in nonequilibrium thermodynamics of classical [5-17] and quantum systems [18-21], and experimentally realized Brownian and electronic systems [22-27], set new bounds on the demon's efficiency. And the question as to whether these bounds are sharp and how they can be realized in experiment is still open.

Here, we examine a bound on demons, i.e. information engines, that follows from a generalization of Jarzynski equality [28] to feedback-controlled systems [8,9,15,17,29],

$$\left\langle \exp\left[-\beta(W-\Delta F)-(I-I_u)\right]\right\rangle = 1. \quad (1)$$

The exponent averaged in Eq. (1) augments the terms from the standard Jarzynski equality – the work performed on the system $W$ and the free energy change $\Delta F$ (in $k_B T = \beta^{-1}$ units) – with a contribution from the information circuitry: $I$ is the information gathered by measurements, out of which a part $I_u$ becomes unavailable due to the irreversibility of the feedback process [17]. Applying Jansen's inequality to Eq. (1) yields a generalized second law [17],

$$\langle -\beta W \rangle \leq -\beta \Delta F + \langle I \rangle - \langle I_u \rangle. \quad (2)$$

Namely, the work extracted from the information engine $\langle -\beta W \rangle$ (in $k_B T = \beta^{-1}$ units) cannot exceed the sum of the free energy difference between final and initial states $-\beta \Delta F$ and the available information $\langle I \rangle - \langle I_u \rangle$. In the absence of information, the inequality recaps the notion of the free energy as the maximal available work in an isothermal process, $\langle -W \rangle \leq -\Delta F$, while the additional term $\langle I \rangle - \langle I_u \rangle$ sets an upper bound on extra work that can be gained from information on the system. We call an information engine "*lossless*" if it achieves the tight bound of Eq. (2). This indicates that almost none of the available information from the feedback protocol is lost, while it does not exclude other, more energetically efficient protocols. We also note that the derivation of Eqs. (1) and (2) does not account for the external energetic cost of detecting the particle and moving the trap accordingly [17].

In this paper, we use an information engine made of a colloidal particle trapped in a harmonic potential to demonstrate the sharpness of the bound set by the generalized second law in Eq. (2). During each cycle of the engine, a high-precision measurement of the particle position is followed by a swift shift of the trap according to the measurement and thermal relaxation of the particle before the next cycle begins. Iterating the measurement-feedback-relaxation cycle, the engine can transport the particle unidirectionally, thereby extracting work from the random thermal fluctuations of the surrounding heat bath. We derive the optimal operating point of the engine when the work extracted per cycle peaks, and show that this peak reaches the bound in Eq. (2). We also show that the engine satisfies the generalized Jarzynski equality in Eq. (1). Thus, we validate these basic nonequilibrium bounds in a nearly error-free feedback control system.

*The Brownian information engine.* – We investigate a simple information engine with one degree-of-freedom, $x_B$, the position of a Brownian particle immersed in a heat bath of temperature $T$ (Fig. 1). The experimental setup is detailed in the following, but first we discuss the basic physical features of this information engine in terms of a simplified model. We consider a particle trapped in the harmonic potential generated by optical tweezers, $V(x_B) = (k/2)(x_B - x_0)^2$, where $x_0$ is the center of the trap and $k$ its stiffness. In the low-Reynolds regime, the dynamics of the particle is overdamped [30,31], with a relaxation time $\tau = \gamma / k = 6\pi \eta R / k \sim 3$ ms, where $\gamma$ is the Stokes friction coefficient. Each cycle consists of measurement, feedback control and relaxation (Fig. 1). The cycle begins when the particle is at thermal equilibrium with a Boltzmann distribution, $P(x) = (2\pi\sigma^2)^{-1/2} \exp[-x^2/(2\sigma^2)]$, where $x = x_B - x_0$ is the deviation from the center $x_0$ with a variance $\sigma^2 = k_B T / k$. A nearly error-free measurement of $x$ is then taken and serves as an input for the following feedback response: we define a region $R$ from $x_0 + L$ to infinity ($L > 0$). Whenever the particle is found in $R$, we instantaneously (i.e. much faster than $\tau \sim 3$ ms) shift the potential center $2L$ to the right; otherwise, if the particle is outside $R$, the potential remains centered at $x_0$. After the feedback step, we let the particle fully relax before we reiterate the cycle.

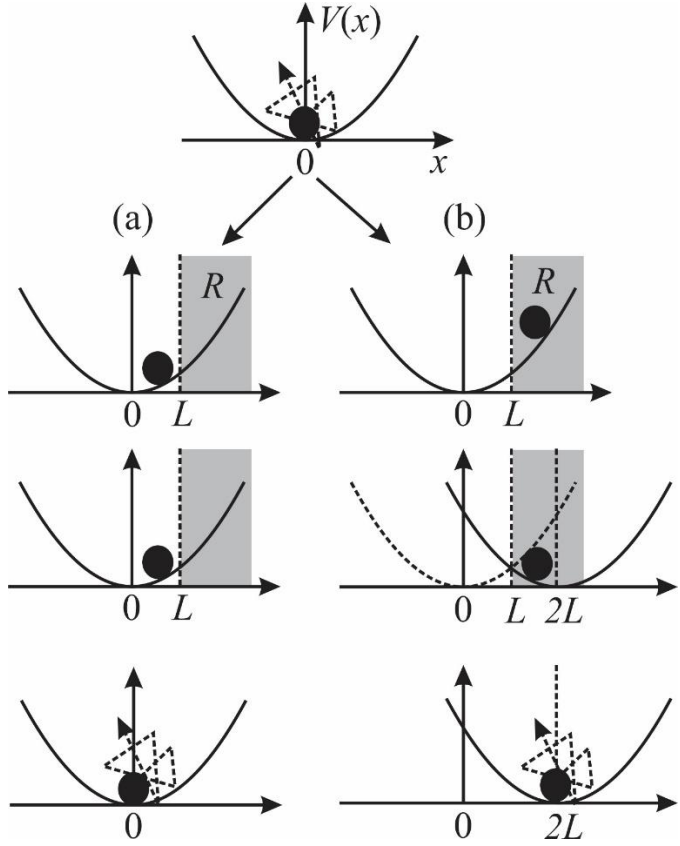

FIG. 1. The measurement-feedback-relaxation cycle of the Brownian information engine. A particle is initially in thermal equilibrium in a harmonic potential $V(x_B) = (k/2)(x_B - x_0)^2$ generated by an optical trap. The feedback is determined as following: We set a region $R$ from $L$ to infinity (shaded). (a) If the particle is outside $R$, nothing is changes. (b) If the particle is inside $R$, we instantaneously shift the potential center to $2L$. By shifting the potential center, the system extracts work equal to the change in the potential energy $\Delta V$. After the feedback step, the system relaxes back to thermal equilibrium and the cycle repeats.

Next, we consider the energy balance during a shift cycle. After the shift, the particle always returns to the same equilibrium macrostate and the free energy remains unchanged, $\Delta F = 0$. Moreover, in the overdamped regime we can disregard the kinetic energy of the particle, so the change in the potential energy when the trap shifts, $\Delta V(x)$, is fully converted into heat and work [32]. However, the potential is shifted much faster than the typical relaxation time such that the particle has no time to move and dissipate energy [33]. Therefore, all the potential energy gained by the shift is converted into work. During the relaxation step, no work is done and only heat is produced by thermal dissipation. We conclude that the extractable work is $-W(x) = -\Delta V(x) = V(x) - V(x-2L) \geq 0$ when the trap shifts ($x \geq L$), and $-W(x) = 0$ otherwise ($x \leq L$). The average work extracted is therefore

$$\langle -\beta W \rangle = -\int_{-\infty}^{\infty} dx\, P(x) \beta W(x) = 2\pi^{-1/2} \ell e^{-\ell^2} - 2\ell^2\, \text{erfc}(\ell), \tag{3}$$

where $\ell = 2^{-1/2}(L/\sigma)$ and $\text{erfc}(z) = 2\pi^{-1/2} \int_z^{\infty} e^{-t^2} dt$ is the complementary error function. Since $\langle -W \rangle > 0$ for any positive $\ell$, the feedback mechanism always allows to extract work from the system.

To examine whether our feedback protocol can achieve the upper bound on the extractable work, we evaluate the terms in Eq. (2). Since the measurement is practically error-free, the net information is simply Shannon's entropy of a Gaussian variable [17,34]:

$$\langle I \rangle = \langle -\ln[P(x)\Delta] \rangle = -\lim_{\Delta \to 0} \int_{-\infty}^{\infty} dx\, P(x) \ln[P(x)\Delta] = \lim_{\Delta \to 0} \ln \sqrt{2\pi e (\sigma/\Delta)^2}, \tag{4}$$

where the limit of vanishing measurement error $\Delta \to 0$ ensures the positive-definiteness of the entropy and the correspondence between discrete and differential entropies ([35] Ch. 8). During the relaxation phase of the feedback process, part of the information in Eq. (4) becomes unavailable [17]. To calculate the unavailable information $\langle I_u \rangle$ we consider the inverse process: the particle is initially in equilibrium with the center of the trap at $x_0 + 2L$ and we shift the potential center back to $x_0$ according to the same protocol. The unavailable information associated with a single measurement is $I_u = -\ln[P(x)\Delta]$ for $-\infty \leq x \leq L$, and $I_u = -\ln[P(x-2L)\Delta]$ for $L \leq x \leq \infty$. The average unavailable information is therefore,

$$\langle I_u \rangle = -\lim_{\Delta \to 0} \left\{ \int_{-\infty}^{L} dx\, P(x) \ln[P(x)\Delta] + \int_{L}^{\infty} dx\, P(x) \ln[P(x-2L)\Delta] \right\}$$
$$= -2\pi^{-1/2} \ell e^{-\ell^2} + 2\ell^2\, \text{erfc}(\ell) + \lim_{\Delta \to 0} \ln \sqrt{2\pi e (\sigma/\Delta)^2}. \tag{5}$$

The upper bound of extractable work is found from Eqs. (2), (4) and (5) keeping in mind that $\Delta F = 0$,

$$\langle -\beta W \rangle \leq \langle I \rangle - \langle I_u \rangle = 2\pi^{-1/2} \ell e^{-\ell^2} - 2\ell^2\, \text{erfc}(\ell). \tag{6}$$

Comparing Eqs. (3) and (6), we see that the present feedback protocol achieves the equality in the generalized second law in Eq. (2) for any $L$, $\langle -\beta W \rangle = \langle I \rangle - \langle I_u \rangle$, and is therefore lossless. This is because the feedback is instantaneous and $W(x)$ has no time to fluctuate. In the following, we experimentally confirm the equality in Eq. (6) by measuring the average work $\langle -\beta W \rangle$ and comparing it to the available information $\langle I \rangle - \langle I_u \rangle$ (r.h.s. in Eq. (6)). Finally, we verify that the feedback protocol satisfies the generalized Jarzynski equality in Eq. (1) by substituting the work and the information terms,

$$\langle e^{-\beta W - I + I_u} \rangle = \int_{-\infty}^{L} dx P(x) + \int_{L}^{\infty} dx P(x) e^{-\beta[V(x-2L)-V(x)]} \frac{P(x)}{P(x-2L)} = 1. \quad (7)$$

*Experimental setup.* – The schematic of our home-built optical tweezers set up is shown in supplementary information as Fig. S1. A laser with 1064 nm wavelength is used for trapping a colloidal particle. The laser is fed to the Acousto-Optic Deflector (AOD) (Isomet, LS110A-XY). The AOD is controlled via an analog voltage controlled Radio-Frequency (RF) synthesizer driver (Isomet, D331-BS) and is capable of diffracting the laser light. The first order diffracted beam is focused at the sample plane of an optical microscope (Olympus IX73) using a 100x oil immersion objective lens of 1.30 numerical aperture. A second laser with 980 nm wavelength is used for tracking the particle position. A Quadrant Photo Diode (QPD; S5980, Hamamatsu) is used to detect the particle position. The electrical signal from QPD is pre-amplified by the signal amplifier (OT-301, On-Trak Photonics, Inc.) and sampled periodically with a Field-Programmable Gate Array (FPGA) data acquisition card (National Instruments, PCI-7830R). The QPD is capable of tracking the particle position with high spatial accuracy of 1 nm [36]. This is sufficiently enough to assume that our system is capable of performing nearly error-free measurements. We have designed a real-time feedback control system using LabVIEW programmed on the FPGA target. The feedback control measurement system is capable of position detection, potential modulation, and data storage. The sample cell consists of the highly dilute solution of 1.99 µm diameter polystyrene particles suspended in deionized water. The trapping laser power at the sample stage is maintained at ~3 mW. Whereas, the laser power of the tracking laser is fixed at ~5% of the trapping laser power. All experiments were carried out at a constant temperature of 300 ± 0.1 K.

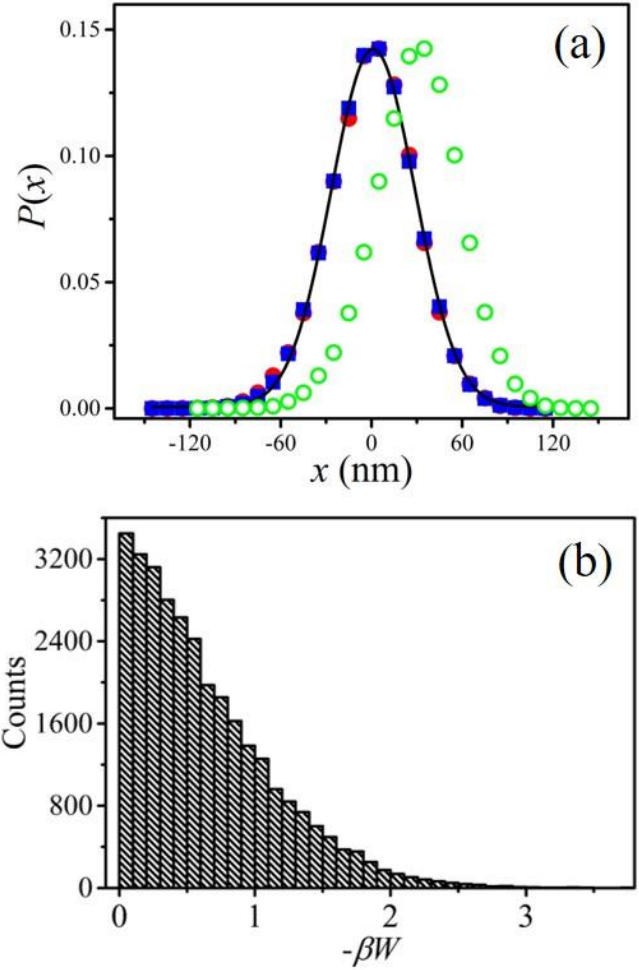

FIG. 2 (color online). (a) Probability distribution $P(x)$ of the particle position in thermal equilibrium (red solid circles), and measured every 25 ms during each engine cycle (blue solid squares) for $L = 0.5\sigma$. The shifted probability distribution $P(x-2L)$ was obtained by performing the backward protocol (green open circles). The black solid curve is a fit to the Boltzmann distribution, $P(x) \propto \exp[-x^2/(2\sigma^2)]$. (b) Histogram of the extracted work $-\beta W$ for $L = 0.5\sigma$.

*Experimental testing of the information engine bounds.* – We first calibrate the parameters of the trap (Fig. 2). By fitting the probability distribution of the particle position in thermal equilibrium without feedback process to the Boltzmann distribution $P(x) = (2\pi\sigma^2)^{-1/2}\exp[-x^2/(2\sigma^2)]$, we obtain the standard deviation $\sigma = 28$ nm. The trap stiffness was calibrated using two different techniques based on the equipartition theorem and the Boltzmann distribution [36], which together give an estimate of $k = 5.28$ pN/μm. Prior to actual operating of the information engine, we set the region $R$ from $L > 0$ to infinity. The QPD measures the particle position periodically at intervals of 25 ms. The FPGA board generates a bias voltage that corresponds to the initial position of the potential center. This bias voltage is applied to the AOD via the RF synthesizer driver. If the particle is found in $R$, the FPGA board generates an updated bias voltage that corresponds to the shift of the potential center to $2L$. The decision whether to update the bias voltage and thereby shift of the potential center is taken within 20 μs. After shifting the potential center, we wait for 25 ms, about eight times the relaxation time $\tau \sim 3$ ms. Finally, the potential center is instantaneously shifted back (within ~20 μs) to the initial position, we wait for 25 ms for full relaxation of the particle and the cycle is repeated.

We next focus on the energetics of the information engine. We set the region $R$ from $L = 0.5\sigma = 14$ nm to infinity and perform the measurement and feedback control described above. The distribution (blue squares in Fig. 2(a)), which is obtained from 100,000 feedback cycles, is indistinguishable from the equilibrium distribution (red) with the same $\sigma = 28$ nm. Fig. 2(b) shows the distribution of the measured extracted work, $-\beta W(x) = \beta[V(x) - V(x-2L)]$ for $x \geq L$ and $-\beta W(x) = 0$ for $x \leq L$, whose average is $\langle -\beta W_{\exp} \rangle = 0.197 \pm 0.001$. We also calculated the average extractable work from the model in Eq. (3), $\langle -\beta W_{\text{model}} \rangle = 0.198 \pm 0.002$, which agrees well with the experimental value. This shows that the feedback protocol is capable of extracting positive work from the information of the system immersed in a single heat bath, thus exceeding the standard bound of the second law of thermodynamics ($\langle -W \rangle \leq -\Delta F = 0$).

Using the equilibrium probability distribution $P(x)$ and the shifted one $P(x-2L)$ from Fig. 2(a), we measured available information $\langle I \rangle - \langle I_u \rangle$ from definitions in Eqs. (4) and (5), where integration was approximated by discrete summation. The experimental value of the available information $\langle I \rangle - \langle I_u \rangle = 0.200 \pm 0.002$, is close to the measured extracted work. This demonstrates the sharpness of the bound set by the generalized second law with an efficiency of information-to-energy conversion of $\langle -\beta W \rangle / (\langle I \rangle - \langle I_u \rangle) = 98.5 \pm 1.1\%$. To find the optimal feedback protocol, we evaluated the extracted work as a function of $L/\sigma$ (red solid circles in Fig. 3(a)). The fit to the theoretical curve in Eq. (3) agrees well with the measurement, implying that the engine indeed achieves the upper bound of the generalized second law in Eq. (2) for any $L > 0$. The maximum of Eq. (3) is obtained when $\exp(-\ell^2) = 2\pi^{1/2} \ell \operatorname{erfc}(\ell)$ at $L \approx 0.612\sigma$. Another quantity of interest is the average step per cycle $\langle \Delta x \rangle$, which measures the average

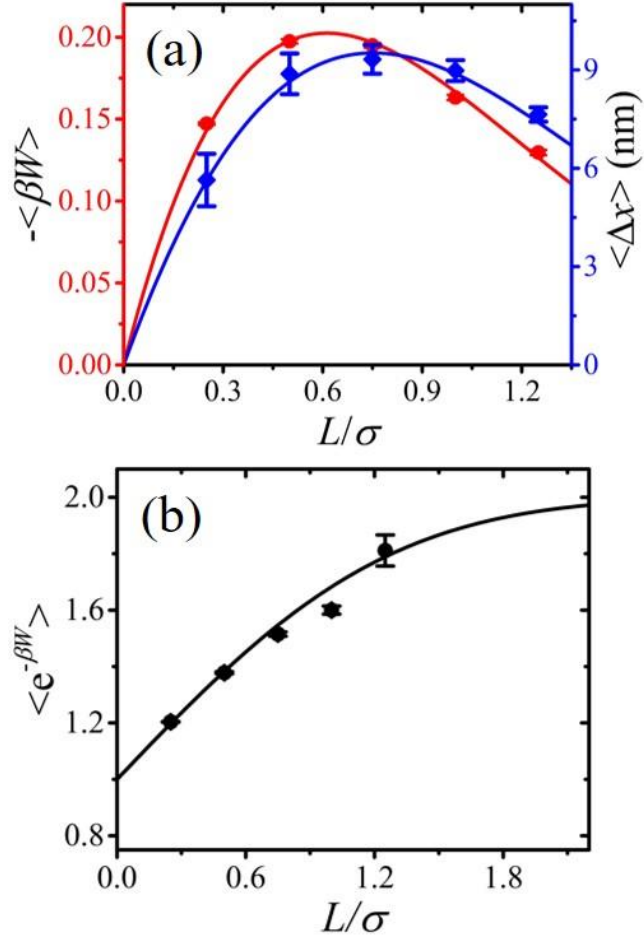

FIG. 3 (color online). (a) Extracted work as a function of $L/\sigma$ (red). Bars denote the standard error of the mean. The red solid curve is a fit to Eq. (3). Plot of average step per cycle (blue). The blue solid curve is a fit to Eq. (8). (b) Plot of $\langle \exp(-\beta W) \rangle$ as a function of $L/\sigma$. The solid curve is obtained by fitting the data with $\mathrm{erfc}(-\ell)$.

unidirectional motion of the Brownian particle,

$$\langle \Delta x \rangle = 2L \int_L^\infty dx P(x) = 2^{1/2} \sigma \ell \, \mathrm{erfc}(\ell). \tag{8}$$

The average step exhibits a maximum of $\langle \Delta x \rangle = 0.34\sigma$ at $L = 0.75\sigma$. The blue solid diamonds in Fig. 3(a) shows the plot of $\langle \Delta x \rangle$ as a function of $L/\sigma$ fits well with Eq. (8). This demonstrates that the optimal working points of maximal extracted work and maximal step are rather close to each other.

We also demonstrated that the generalized Jarzynski equality in Eq. (1) is satisfied by the feedback protocol. To this end, we evaluated the integration in Eq. (7) for $L = 0.5\sigma$ and found it to be equal to 1.02 ± 0.06 (see Fig. S2 in supplementary information) as in Eq. (1). Another formulation of the generalized Jarzynski equality is in terms of the efficacy $\gamma$, defined as the sum of probabilities that time-reversed trajectories are observed,

$$\langle \exp(-\beta W) \rangle = \gamma = \int_{-\infty}^{L} dx P(x) + \int_{L}^{\infty} dx P(x-2L) = \mathrm{erfc}(-\ell). \qquad (9)$$

To verify Eq. (9), we measured the value of $\langle \exp(-\beta W_{\exp}) \rangle = 1.378 \pm 0.004$. We also measured the efficacy parameter from the integral in Eq. (9) and found $\gamma_{\exp} = 1.390 \pm 0.003$. Similarly, we calculated the value of $\mathrm{erfc}(-\ell) = 1.383 \pm 0.013$. All three values agree well with each other, implying that the generalized Jarzynski equality in Eq. (9) is satisfied. Fig. 3(b) shows a plot of $\langle \exp(-\beta W_{\exp}) \rangle$ as a function of $L/\sigma$, which agrees well with a fit to the model $\mathrm{erfc}(-\ell)$.

In conclusion, we examined a simple information engine consisting of a colloidal particle trapped by optical tweezers. By precisely measuring the particle position and shifting the potential center practically instantaneously, we can extract positive work from a system in a single heat bath at a constant temperature, thus exceeding the conventional bound of the second law of thermodynamics. The extra work originates from information on the system, which allows the feedback protocol to generate unidirectional motion. The measured work agrees well with the theoretical prediction, and we found that maximum work can be extracted from the engine when $L \approx 0.612\sigma$. Finally, we demonstrated that the feedback protocol satisfies the generalized Jarzynski equality and is able to achieve the equality in the generalized second law under error-free measurements. Hence, the bound on information engines (demons) from Eq. (2) is sharp.

This work was supported by Grant No. IBS-R020-D1by the Korean government and UNIST Research Fund 1.140104 (H. K. P.). We thank Prof. Masahito Ueda for useful discussions.

*Electronic address: hyuk.k.pak@gmail.com

†Present address: Departament de Física de la Matèria Condensada, Universitat de Barcelona, Barcelona 08028, Spain

# Supporting information for

# Lossless Brownian information engine

Govind Paneru,[1] Dong Yun Lee,[1,†] Tsvi Tlusty,[1,2] and Hyuk Kyu Pak[1,2*]

[1]Center for Soft and Living Matter, Institute for Basic Science (IBS), Ulsan 44919, Republic of Korea

[2]Department of Physics, Ulsan National Institute of Science and Technology (UNIST), Ulsan 44919, Republic of Korea

[†]Present address: Departament de Física de la Matèria Condensada, Universitat de Barcelona, Barcelona 08028, Spain

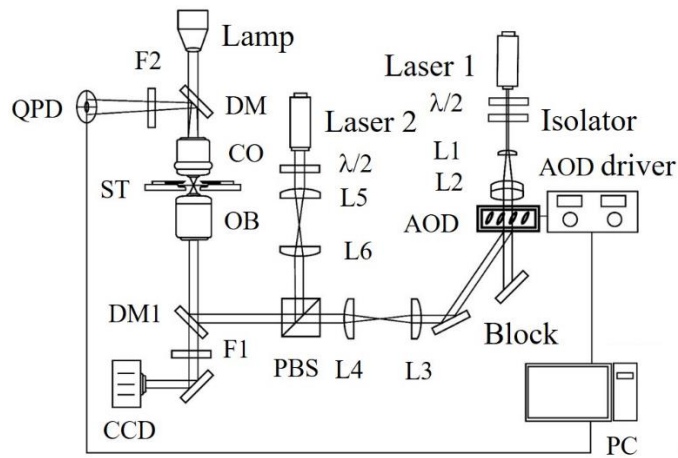

FIG. S1. Schematic of a home built optical tweezers.

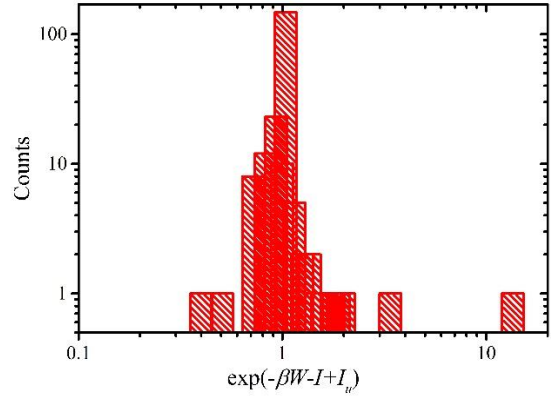

FIG. S2. Distribution of $\exp(-\beta W - I + I_u)$ in order to verify the generalized Jarzynski equality in Eq. (1). The average is $\langle \exp(-\beta W - I + I_u) \rangle = 1.02 \pm 0.06$.